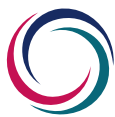

# Counterion effects on nano-confined metal–drug–DNA complexes

Nupur Biswas[1,2], Sreeja Chakraborty[3], Alokmay Datta[*1], Munna Sarkar[3], Mrinmay K. Mukhopadhyay[1], Mrinal K. Bera[4] and Hideki Seto[5]

Address:
[1]Surface Physics and Material Science Division, Saha Institute of Nuclear Physics, 1/AF Bidhannagar, Kolkata 700064, India, [2]present affiliation: Soft Condensed Matter Department, Raman Research Institute, Bangalore 560080, India, [3]Chemical Science Division, Saha Institute of Nuclear Physics, 1/AF Bidhannagar, Kolkata 700064, India, [4]Center for Advanced Radiation Sources, University of Chicago, Chicago, Illinois 60637, USA and [5]KENS & CMRC, Institute of Materials Structure Science, High Energy Accelerator Research Organization, Tsukuba 305-0801, Japan

Email:
Alokmay Datta[*] - alokmay.datta@saha.ac.in

* Corresponding author

## Abstract

We have explored morphology of DNA molecules bound with Cu complexes of piroxicam (a non-steroidal anti-inflammatory drug) molecules under one-dimensional confinement of thin films and have studied the effect of counterions present in a buffer. X-ray reflectivity at and away from the Cu K absorption edge and atomic force microscopy studies reveal that confinement segregates the drug molecules preferentially in a top layer of the DNA film, and counterions enhance this segregation.

## Introduction

Condensed state behaviour of DNA, the best-known biopolymer, in a confined space is a matter of interest due to its relevance in living systems. Within cells DNA molecules remain in a confined space crowded by other molecules and ions. Thus, there are three aspects of the situation, which demand elucidation – the role of the ions, of the molecules (especially macromolecules) and of the confinement in the length scales of nanometers and micrometers – in maintaining the stability and homogeneity of the phase of the mixture [1-4] as well as of the structure of the DNA molecules [5]. Studies on the first aspect have established that depending on counterion concentrations and valencies, DNA molecules in bulk solution exhibit isotropic to liquid crystalline phase transition and under extreme conditions they can form crystalline states [6]. Again, a mixture of DNA and other macromolecules undergoes spontaneous segregation and organization under micrometre-scale confinement [7]. Regarding confinement effects at the nanometer scale, we have observed that in absence of counterions DNA molecules form layered structures aligned laterally to the film surface, whereas in case of films prepared from buffered solution there is no such layering due to the increased orientational entropy of entangled shorter DNA molecules [8,9]. This indicates that both confinement and presence of charged and neutral species in the environment dictates the structure and dynamics of DNA aggregation and lead us to explore the confinement effect on other biologically relevant DNA composites.

We have focussed on the effect of one such biologically active molecule, piroxicam, which is an enolic acid. It is used as a nonsteroidal anti-inflammatory drug (NSAID) for symptomatic relief from rheumatoid arthritis, osteoarthritis and spondylitis [10]. However, the metal-complexes of this molecule form another group of drugs of even greater interest due to their anti-cancer activity [11,12]. In this context, the attachment of these drugs to DNA molecule gains special importance, since this determines their biofunctionality [13,14]. It is already reported that a Cu(II) complex of piroxicam intercalates within the DNA backbone [15-17]. Motivated by these observations, we have studied effect of counterions on the confined state of metal–drug–DNA complexes, because within living systems such complexes exist in the presence of various salt ions, in a highly confined state.

Here we report structural studies of a metal–drug–DNA complex within thin films, i.e., under one-dimensional confinement and of the influence of counterions on this confined system. Specifically, we have studied thin films comprising of composites of DNA and a Cu(II) complex of piroxicam in presence and absence of buffer molecules. Using anomalous X-ray reflectivity we have probed its out-of-plane structure whereas atomic force microscopy has provided us its in-plane morphology.

## Experimental

Polymerized calf thymus DNA (Sisco Research Laboratory, India) dissolved in triple-distilled water formed a pristine stock solution. The absorbance ratio $A_{260}/A_{280}$ of the solution at 260 and 280 nm being in the range $1.8 < A_{260}/A_{280} < 1.9$, indicated that no further deproteinization of the solution was necessary. The nucleotide concentration of the stock solution, assuming $\varepsilon_{260} = 6600$ $M^{-1}cm^{-1}$, was found to be 1.8 mM. The stock solution was diluted to the desired concentration of 800 μM in triple distilled water. 10 mM of sodium cacodylate (Merck, Germany) solution in triple-distilled water was adjusted to the desired pH of 6.7 with hydrochloric acid and was used as stock buffer solution. Each molecule of sodium cacodylate effectively contributes one $Na^+$ ion.

The Cu–piroxicam complex was synthesized following an established protocol [18] and was characterized by FTIR, EPR,

and elemental analysis as described in a previous work [19]. Figure 1 shows the structure of the piroxicam molecule along with its Cu(II) complex. Piroxicam was purchased from Sigma Chemicals and was used without purification. A stock solution of piroxicam of strength 0.5 mM was prepared in spectroscopic grade ethanol, because of poor solubility of piroxicam in water. CuCl₂ was purchased from Sigma Chemicals and stock solution was prepared by dissolving it in water. Concentration of stock Cu(II)–piroxicam solution was maintained at 37.4 μM. This stock solution was mixed with DNA stock solution of concentration 800 μM resulting in a fraction of 0.046 of drug–metal complex in DNA. We have worked with two solutions of drug–metal complexes mixed with DNA, one in presence and the other in absence of buffer. When mixed with DNA-buffer solution, the buffer concentration was maintained at 10 mM.

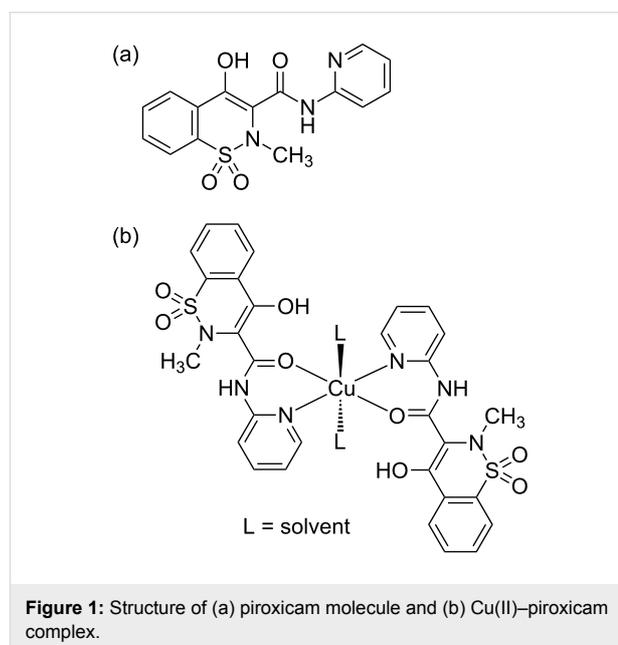

**Figure 1:** Structure of (a) piroxicam molecule and (b) Cu(II)–piroxicam complex.

Films were prepared by spin-coating the solution on amorphous fused quartz substrates at ambient condition using a spin-coater (Headway Research Inc., USA). Before spin-coating the fused quartz (Alfa Aesar, USA) substrates were cleaned and hydrophilized by boiling in 5:1:1 $H_2O/H_2O_2/NH_4OH$ solution for 10 min, followed by sonication in acetone and ethanol respectively, then rinsing by Millipore water (resistivity ≈ 18.2 MΩ·cm) and subsequent removal of water by spinning the substrate at high speed (4000 rpm).

To extract out-of-plane information specular X-ray reflectivity profiles of these thin films were recorded with step size 5 mdeg at the Indian Beamline (BL-18B) at Photon Factory, High Energy Accelerator Research Organization (KEK), Japan. Both anomalous and normal X-ray reflectivity data was collected to check the spatial distribution of Cu along the film depth. Cu has two absorption edges in X-ray regime which are at 8.98 keV (K1 edge) and 1.096 keV (L1 edge). As Indian Beamline operates within 6–20 keV, anomalous reflectivity data was taken at K1 absorption edge of Cu (at wavelength 1.38 Å, energy 8.98 keV). For normal X-ray reflectivity we chose a wavelength which is away from the absorption edge. So we shifted to energy 11.736 keV which corresponds to wavelength 1.08 Å. To avoid radiation damage the sample was kept in nitrogen atmosphere. To analyse all XRR data we have used Minpack fitting package [20] based on Levenberg–Marquardt algorithm. It provides the local minimum of nonlinear least squares functions of several variables. Here an iterative process continues and termination occurs when the relative error between two consecutive iterates is below $10^{-9}$. Atomic force microscope (AFM) images recorded in tapping mode using Nanonics Multi-View1000 with glass tips of about 20 nm diameter, provides in-plane information. The images were analyzed using WSxM software [21].

## Results and Discussion

Figure 2 shows the topography of the film surfaces as obtained from AFM. The top surface of the film comprising drug–metal–DNA complex has a lower height variation (rms roughness 4.28 Å) whereas the film with added buffer has an increased roughness (rms roughness 8.97 Å). We took several images from different regions of the sample. Here we are reporting the average roughness values from five such different images. For Cu(II)–piroxicam–DNA film rms roughness is 4.51 Å with a standard deviation of 0.41 Å. For Cu(II)–piroxicam–DNA–buffer film rms roughness is 8.35 Å with a relatively high standard deviation of 1.08 Å.

This is also apparent from the typical line profiles shown in Figure 2. The phase images shown inset indicate the clear presence of clusters of a material different from the film at the base of the buffered film, whereas such clustering is not so pronounced in the unbuffered film. We have carried out X-ray reflectivity (XRR) experiments of the films, which provide us electron density profiles (EDP) of the film along its depth. To analyze the anomalous reflectivity profiles we have used the distorted wave Born approximation (DWBA) method [22,23], which only requires an ansatz of the average electron density of the film and provides the electron densities of different "layers" of the film (of thickness decided by the spatial resolution) through Fourier transforms [24]. In contrast to the usual anomalous scattering analysis formalism [25] here we have not considered any interfacial width between the "layers". The reflectivity profiles of the films and EDPs along the depth of the film extracted from the fits are shown in Figure 3. It is to be mentioned that in the DWBA model "layers" do not extend into the substrate. To plot EDP, as shown in Figure 3b and Figure 3d, we have further convoluted the DWBA EDP with the average electron density of the substrate and of air ($\rho_s = 0.68$ and $\rho_{air} = 0$) and as well as with the roughnesses of those interfaces ($\sigma_{af} \approx 10$ Å and $\sigma_{fs} \approx 7$ Å). The values of these parameters were obtained from the DWBA fit and during fit it was always taken care that the roughness values do not exceed the "layer" thickness.

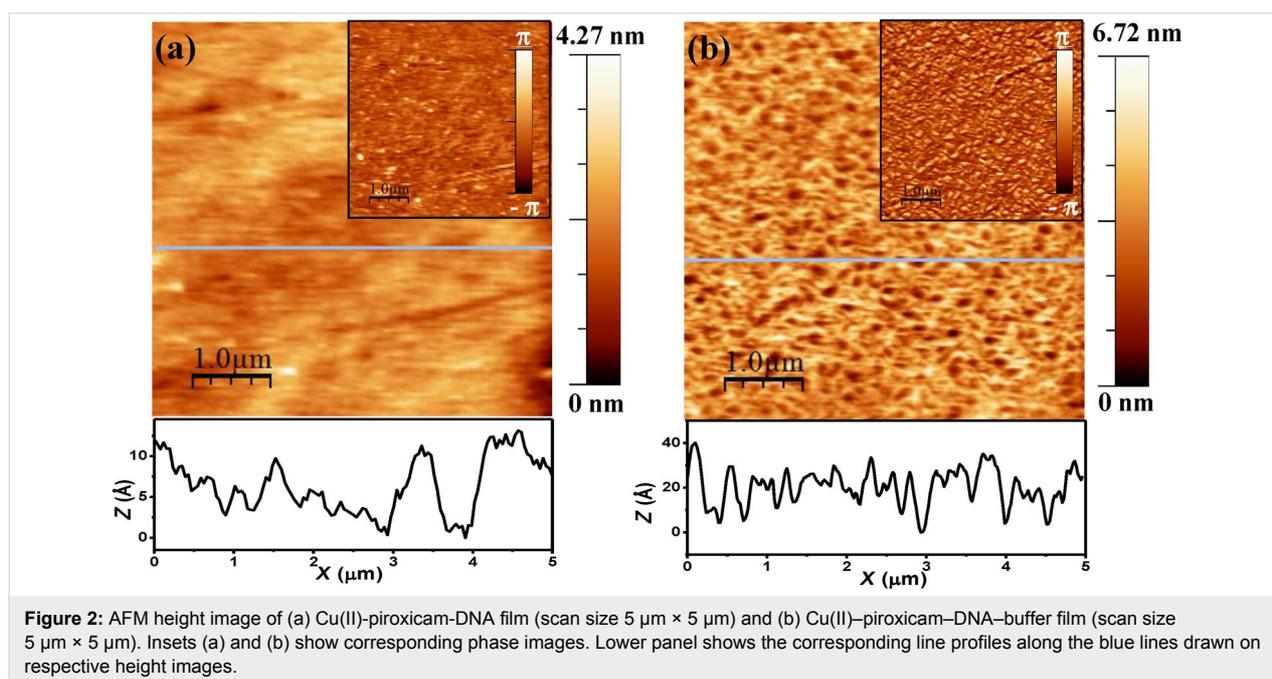

**Figure 2:** AFM height image of (a) Cu(II)-piroxicam-DNA film (scan size 5 µm × 5 µm) and (b) Cu(II)–piroxicam–DNA–buffer film (scan size 5 µm × 5 µm). Insets (a) and (b) show corresponding phase images. Lower panel shows the corresponding line profiles along the blue lines drawn on respective height images.

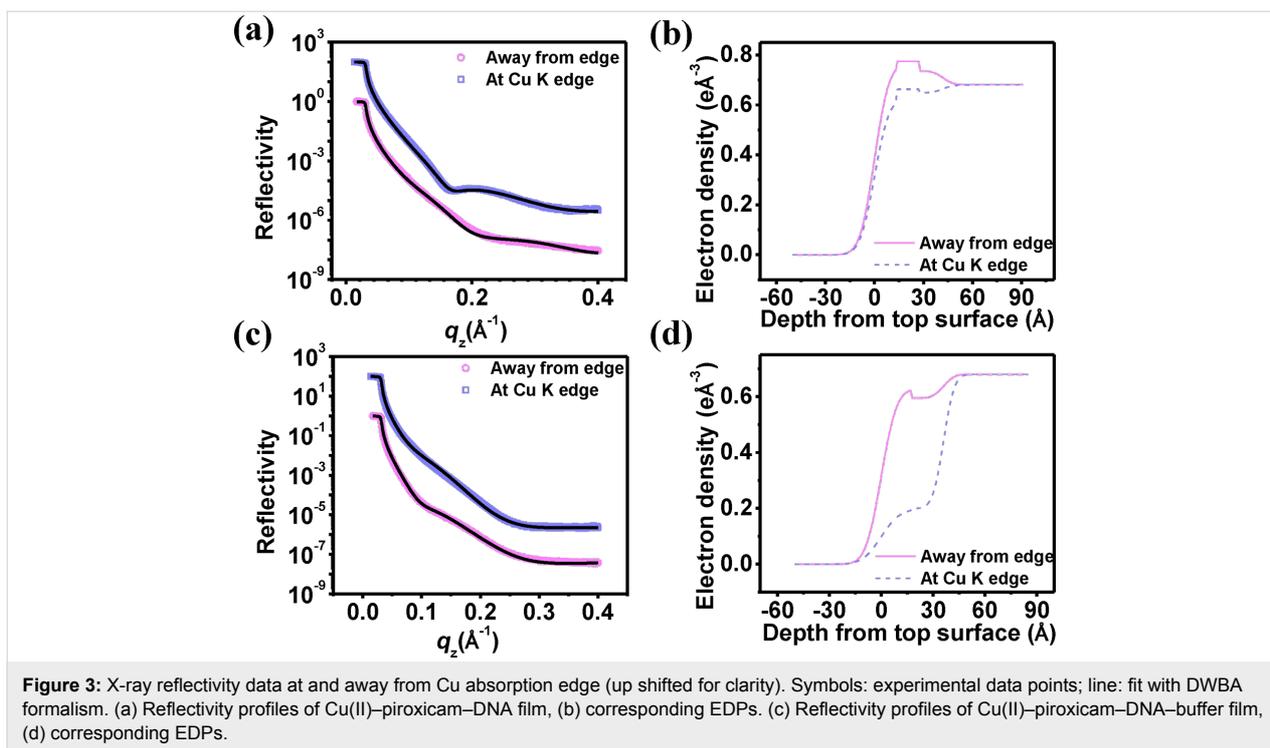

**Figure 3:** X-ray reflectivity data at and away from Cu absorption edge (up shifted for clarity). Symbols: experimental data points; line: fit with DWBA formalism. (a) Reflectivity profiles of Cu(II)–piroxicam–DNA film, (b) corresponding EDPs. (c) Reflectivity profiles of Cu(II)–piroxicam–DNA–buffer film, (d) corresponding EDPs.

We observe that the film prepared from a solution containing no counterions has a thickness of 42 Å, whereas the film prepared from solution with added buffer has a lower thickness of 36 Å. These small thickness values establish that we have succeeded in creating a confined state of this complex. A reduction of the film thickness after addition of buffer can be explained through a better neutralization of DNA molecules by the buffer molecules. This in turn reduces the persistence length of the DNA molecules, makes them softer and more entangled and hence more compact [26]. Considering the fact that DNA molecules have a diameter of 22–26 Å, the magnitude of the film thicknesses suggests a lateral alignment of DNA molecules within the film similar to the case of film formed from "pristine" DNA molecules [8].

Anomalous X-ray scattering data provides us the distribution of an effective density of Cu atoms along the film depth. Within any material X-rays interact with electrons only. Hence, the presence of a particular element is observed by X-ray, when the X-ray energy matches with any absorption edge of that element and the radiation is absorbed. As the beam is no more scattered by the the electrons of that element, the scattered beam provides a lower value of electron density. The effective electron densities for X-ray energies away from edge ($\rho^a(z)$) and at the edge ($\rho^e(z)$) of an element, Cu in our case, is given by [27],

$$\rho^a(z) = N_{Cu}(z) Z^a_{eff} \tag{1}$$

and

$$\rho^e(z) = N_{Cu}(z) Z^e_{eff}, \tag{2}$$

where $N_{Cu}(z)$ is the atomic density of the element Cu. It is independent of energy and effective atomic number $Z_{eff}$ of the element as observed by X-ray. Hence, the difference yields the effective electron density for the element Cu,

$$\rho^a(z) - \rho^e(z) = N_{Cu}(z) \Delta Z_{eff} = \Delta\rho^{eff}_{Cu}(z), \tag{3}$$

where $\Delta Z_{eff} = Z^a_{eff} - Z^e_{eff}$. The difference in electron densities, $\Delta\rho^{eff}_{Cu}$ represents the abundance of Cu. Figure 4a shows this variation along the film depth for both films. We observe a relative abundance of Cu near the air–film interface suggesting a Cu-rich upper layer for both films. This effect is more enhanced in the case of the buffer film, indicating counterions enhance the Cu proportion in this layer. This suggests that a drug–DNA segregation occurs with buffering leaving more drug–metal composites at the surface increasing its roughness and clustering as observed from AFM topography (Figure 2). In Figure 4b we have compared the line profiles drawn over two images. The horizontal lines denote corresponding average height of the profiles (6.38 Å and 20.37 Å for without buffer and with buffer films, respectively). Their difference of approx. 14 Å matches closely with the lateral width of piroxicam molecules [28,29].

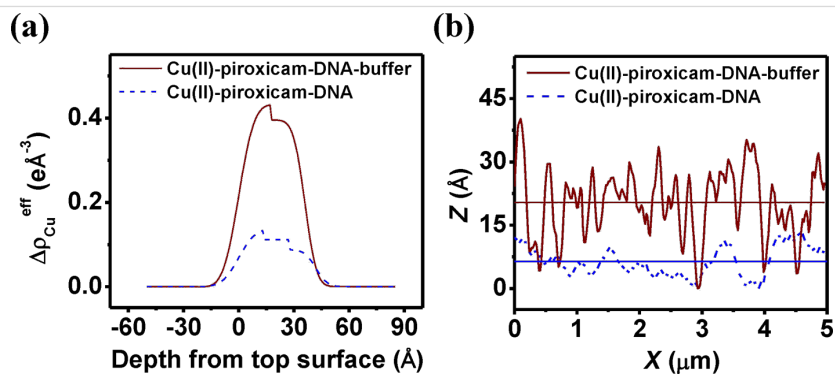

**Figure 4:** (a) Variation of $\Delta\rho_{Cu}^{eff}$ along the film depth for both Cu(II)–piroxicam–DNA and Cu(II)–piroxicam–DNA–buffer films. (b) Comparison of the line profiles along the blue lines drawn on respective height images of Figure 2. The horizontal lines denote corresponding average heights of the profiles.

The structures of piroxicam molecule and Cu(II)–piroxicam complex as depicted in Figure 1 show a planar structure of the complex which intercalates in a DNA backbone parallel to the bases in solution [28]. In our case we observe an asymmetric distribution of Cu atoms with respect to the DNA molecules aligned laterally over the hydrophilic substrate leading to a preferential enrichment at the top surface of the films. It was observed earlier [8,19,30] that the negatively charged phosphate groups of DNA also attach with the hydroxyl-terminated hydrophilic quartz substrate through short-range interactions such as hydrogen bonds that dominating over the long-range but weak, screened Coulomb attraction. Due to this short-range interaction, the hydroxyl-terminated substrate prefers DNA molecules rather than the neutral metal–drug complex. On the other hand, at the surface, due to the absence of any such short-range interactions, the intercalation of the Cu(II) complex of piroxicam is allowed, quite similar to the case of bulk solution. This is shown in Figure 5a. In presence of counterions, the phosphate groups of the DNA backbone get neutralized. This causes two effects, (a) the short-range interactions become even more dominant causing a stronger adhesion to substrate, (b) the presence of counterions in buffer solution neutralizes the polyan-ionic DNA backbone to some extent. Charge neutralization of the backbone reduces the persistence length of DNA, making it more floppy. A floppy DNA is better accommodated nearer to the surface as shown in Figure 5b. Enhanced segmental flexibility of DNA in presence of buffer promotes not only closer approach to the surface but also helps in orienting the Cu-bearing drug molecule more towards the top. These considerations qualitatively explain the enrichment of the Cu–drug complex at the surface and the reduced thickness of the film made from solution with added buffer. This is shown in Figure 5b.

It is to be noted that we have explored the role of very high salt concentrations (500 mM) in case of pristine DNA thin films, as reported in [9]. At such high salt concentrations, we observed salt crystals over the film. Also at concentrations above 100 mM the persistence length of DNA molecules saturates to approx. 50 nm due to complete neutralization of DNA backbone. At salt concentrations of approx. 10 mM DNA is the onset of full neutralization. To explore counterion effects we have restricted ourselves to this onset regime [31].

## Conclusion
We have carried out a preliminary exploration of the morphological effects of counterions on a metal–drug–DNA complex within a thin film. The metal–drug complex intercalates within the DNA backbone and prefers to remain near the top surface of the film. Addition of buffer molecules results in the presence of more metal–drug composites at the top surface of the film and a reduction of the film thickness. We have explained these findings qualitatively, invoking enhanced short-range drug–DNA and substrate–DNA interactions that are influenced by the buffer.

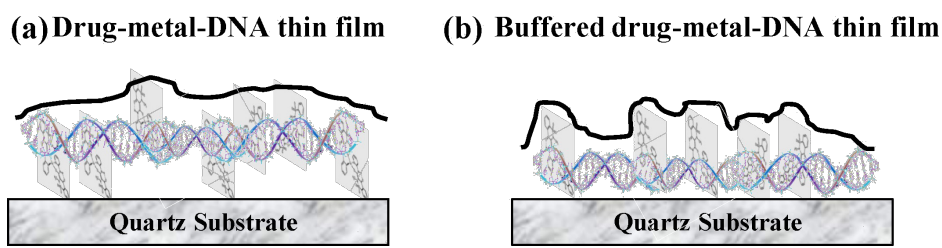

**Figure 5:** Schematic of (a) metal–drug–DNA film and (b) metal–drug–DNA–buffer film.

## Acknowledgements

We would like to acknowledge Heiwa-Nakajima Foundation, Japan for providing financial support and Department of Science and Technology, Government of India for sponsoring Indian beamline project at Photon Factory, KEK, Japan. Authors N. B. and S. C. thank Council of Scientific and Industrial Research (CSIR), Government of India and Director, SINP for their research fellowships.